\documentclass{article}

\usepackage{amsmath,amsfonts,amssymb,amsthm}
\usepackage[dvips]{graphics}
\usepackage{epsfig}

\DeclareGraphicsExtensions{.pdf,.png,.jpg,.tif,.ps}

\theoremstyle{definition}

\begin{document}

\title{Numerical study of the 6-vertex model with domain wall boundary conditions}
\author{David Allison and Nicolai Reshetikhin}

\maketitle

\begin{abstract} A Markov process is constructed to numerically study the
phase separation in the 6-vertex model with domain wall boundary
conditions. It is a random walk on the graph where vertices are
states and edges are elementary moves. It converges to the Gibbs measure of
the 6-vertex model.  Our results show clearly that a droplet of $c$
vertices is created when Boltzamnn weights are in the
antisegnetoelectric region. The droplet is a
diamond-like shaped curve with four cusps.
\end{abstract}

\section*{Introduction}

It is well known  that the 6-vertex model is exactly solvable and
has phase transitions. The history and the classification of phases
in the 6-vertex model as well as many interesting facts about the
structure of the partition function of the 6-vertex model with periodic
boundary conditions can be found in \cite{LWu},
\cite{Bax}.

There is an important function of Boltzmann weights of the model
which is usually denoted by $\Delta$ (see \cite{Bax},\cite{LWu}
and section \ref{results}). The 6-vertex model with periodic 
boundary conditions has 3 phases in the thermodynamical limit,
depending on the value of $\Delta$.
One is the totally ordered (frozen) phase, with $\Delta>1$, the second is the 
disordered (critical) phase, with $|\Delta|<1$, and the third is
the partially ordered (antisegnetoelectic) phase with $\Delta<-1$.

The 6-vertex model with domain wall boundary conditions on a square
$N\times N$ grid perhaps was first
considered in \cite{Kor} in the process of computation of norms of
Bethe vectors. The partiction function of this system can be written
as the determinant of a certain $N\times N$ matrix \cite{Izerg}. 
Its asymptotics in the thermodynamic limit $N\to \infty$ were analyzed in \cite{KoZJ}. It is related to matrix
models, which was pointed out and exploited in \cite{ZJ}.

The 6-vertex model with $\Delta=1/2$ is also  known as the ice-model.
This model with domain wall (DW) boundary condition 
is closely related to the enumeration of alternate sign
matrices \cite{Ku}. It also has other interesting combinatorial
features (see for example \cite{Zub}, \cite{Strog}). When $\Delta=0$
the 6-vertex model is equivalent to the problem of counting of
weighted tilings of the Aztec diamond (see for example
\cite{KoZJ}, \cite{KenDim} and references therein).

The spatial coexistence theory of different phases and
the interfaces separating phases is an important part of
statistical mechanics. Growth of crystals is one of the
well known phenomena of this type. This is also closely
related to the limit shape effect in statistics of
Young diagrams \cite{asymth} and plane partitions.

In dimer models related to enumeration of plane partitions
and domino tilings, the interface between the disordered and totally
ordered phases is also known as an {\it arctic circle} phenomenon \cite{Arctic}.

Dimer models on bipartite planar graphs with
periodic weights are exactly solvable models where this phenomenon
has been studied in \cite{Ken-Okoun-Shef}. In dimer models the limit
shapes or interfaces (curves separating phases), under broad conditions, are real
algebraic curves \cite{Ken-Okoun-Shef}. Since at $\Delta=0$ the
6-vertex model is equivalent to a dimer model these results imply
that such a phenomenon exists in the 6-vertex model for $\Delta=0$.
The natural question is whether the spatial coexistence of phases happens
only at the free fermionic point or if it occurs for all values of
$\Delta$.  Numerical evidence suggesting the existence of a limit
shape in the 6-vertex model with domain wall boundary conditions
for all weights was obtained in \cite{SZ}.

Here we report results of numerical study of the 6-vertex model
with DW boundary conditions in all phases of the model. Our method
is different from \cite{SZ}.  To generate a random configuration
in the 6-vertex model we construct the Markov process which is
equivalent to a random weighted walk on the graph where the
vertices are states of the model and edges are local moves which
transform states into other states. This process satisfies
the detailed balance condition and therefore converges to the
Gibbs state of the 6-vertex model. 
It is also known as Monte-Carlo with local update and as a
heat-bath algorithm. In statistical mechanics such processes
are known as Kawasaki, or Glauber dynamics. For a more effective 
version of this algorithm known as the 
``coupling from the past'' algorithm, see \cite{PW}.

For periodical boundary conditions 
the system can be either in the ordered (segnetoelectric) phase,
disordered (critical) phase, or antisegnetoelectric (non-critical)
phase, depending on the values of Boltzmann weights.

Our results confirm the conclusion from \cite{Er} and \cite{SZ}
that there is a coexistence of ordered and disordered phases in
the 6-vertex model. They also clearly indicate that for
$\Delta<-1$ there is a coexistence of all three phases. The outer
layer is an ordered phase. It follows by the ring of disordered
phase. Finally, there is an inner droplet of the antisegnetoectric
phase. This phenomenon was first conjectured in \cite{SZ} using a
different numerical method.  The shape of the inner droplet has
four cusps and is reminiscent of one of the limit shapes for
dimers on a square-octagon grid \cite{Ken-Okoun-Shef} equivalent
to the diablo tiling .

{\bf Acknowledgments}. We are grateful to  R. Kenyon and A.
Okounkov for many illuminating discussions, to K. Palamarchuk for
valuable comments, and to T. Yates for help with the
implementation of the algorithm in the C language.

\section{Weights and local moves}
\subsection{States}

\begin{figure}[htp]
\begin{center}
\begin{tabular}{cccc}
$a_{1}$ & $b_{1}$ & $c_{1}$\\
\includegraphics[width=1in]{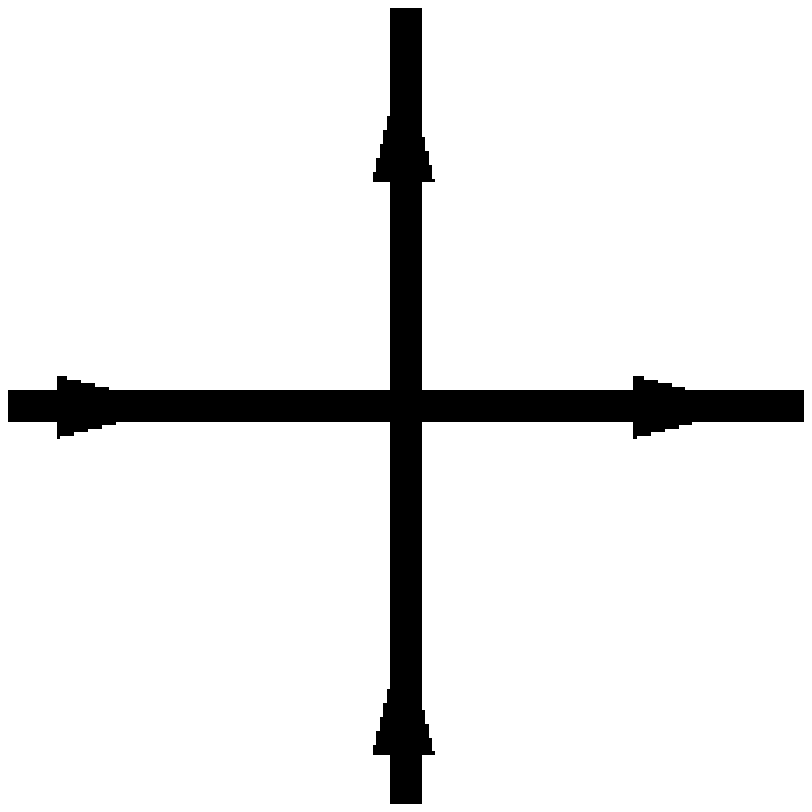} &\includegraphics[width=1in]{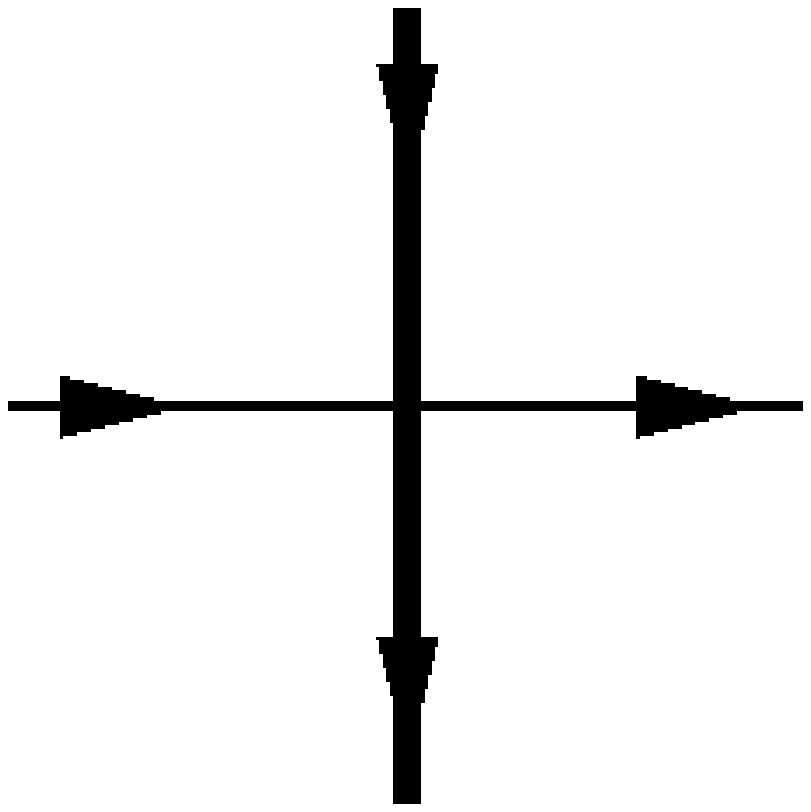} &\includegraphics[width=1in]{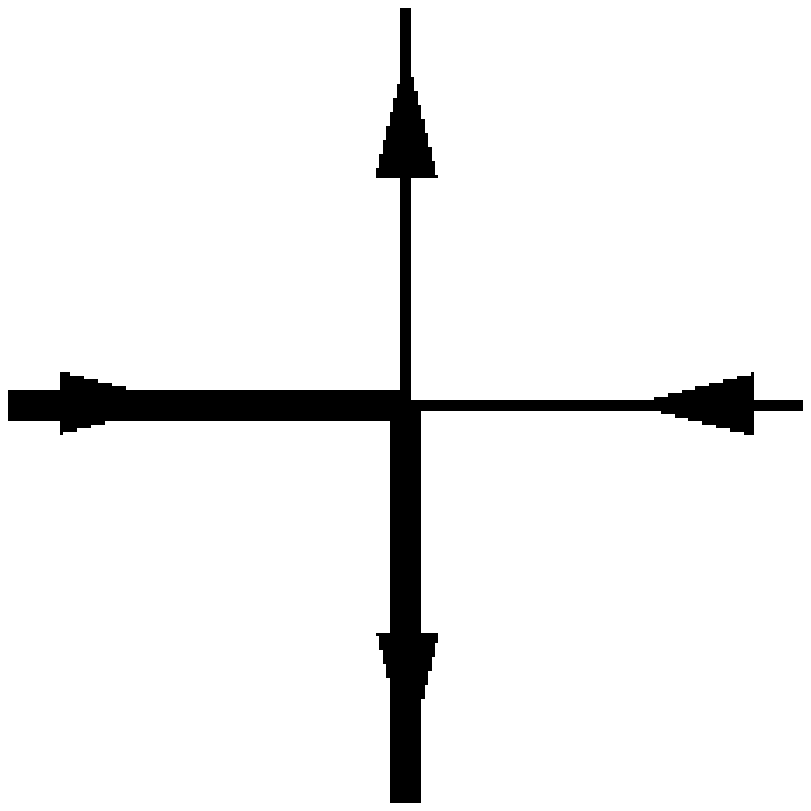}\\
$a_{2}$ & $b_{2}$ & $c_{2}$\\
\includegraphics[width=1in]{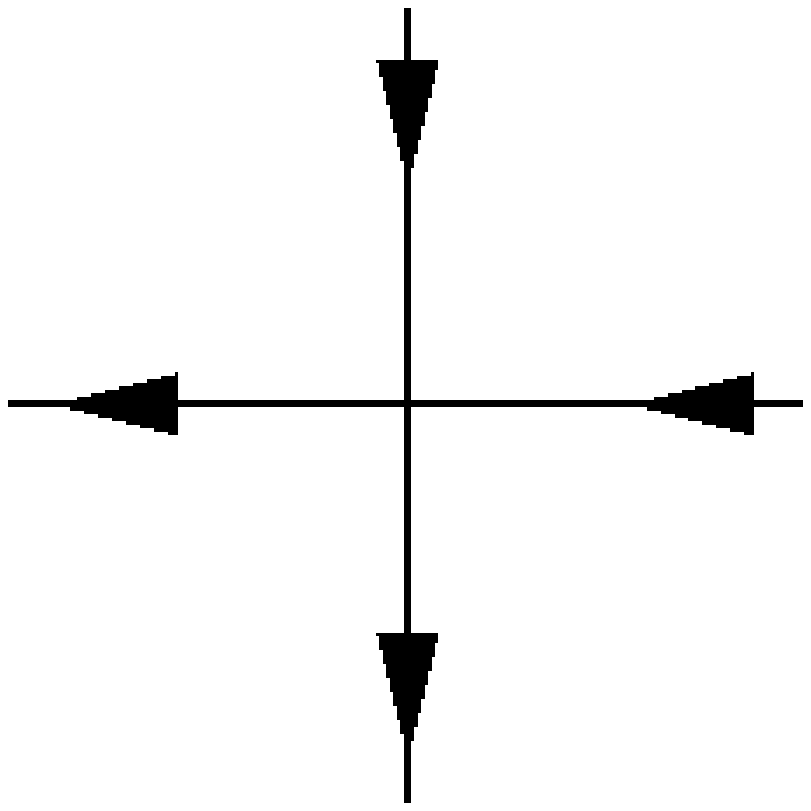} &\includegraphics[width=1in]{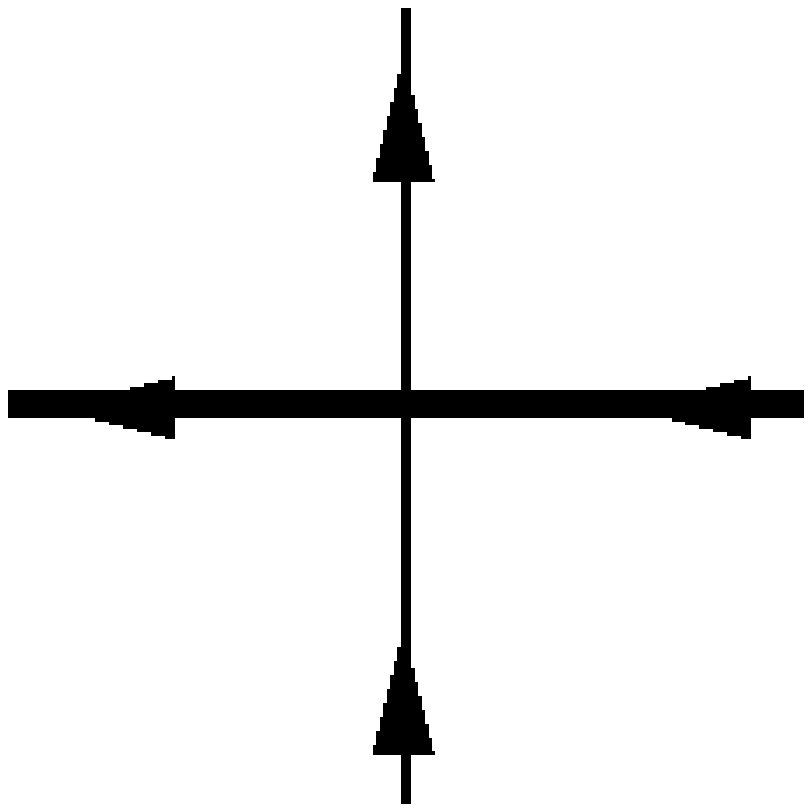} &\includegraphics[width=1in]{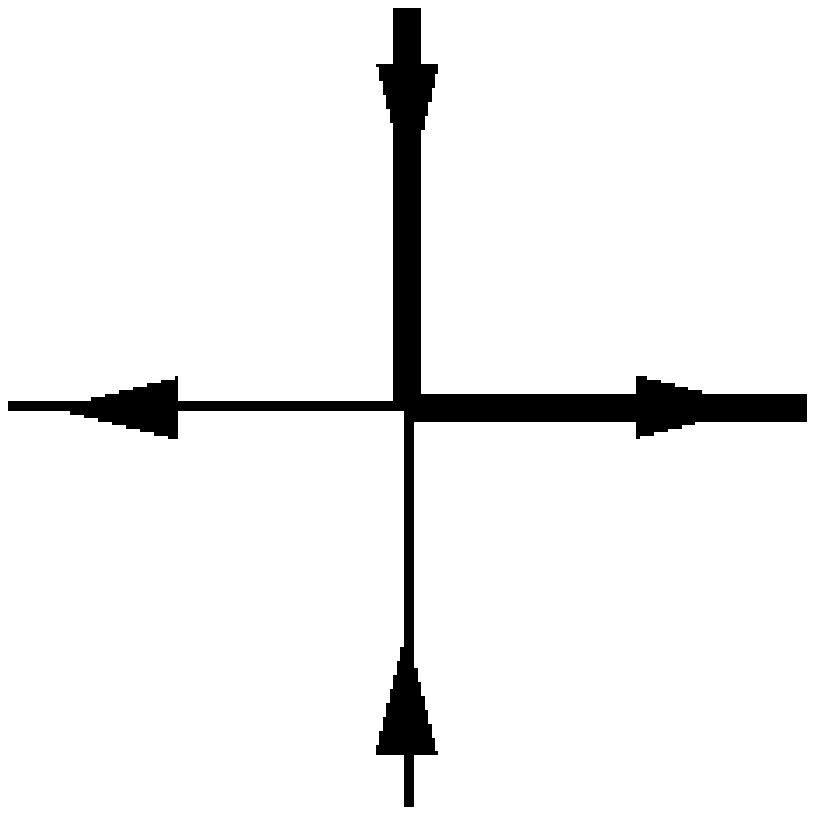}\\
\end{tabular}
\caption{The 6 vertices and their weights}
\end{center}
\label{6-v}
\end{figure}

States of the 6-vertex model on a square lattice are configurations
of arrows assigned to each edge. The 6-vertex rule is that the
total number of arrows coming into any vertex should be equal to
the total number of arrows going out of this vertex.  Each
configuration of arrows can be equivalently regarded as a
configuration of empty edges (arrows oriented South-North and
East-West) and occupied edges, or thick edges (arrows pointing in
the opposite directions). It is clear that thick edges will form
paths. Possible configurations of paths around a vertex are shown
on fig. \ref{6-v}.

We will use $a_1, a_2, b_1, b_2, c_2$, and $c_2$ as names of the
vertices. We denote by the same letters Boltzmann weights assign
to these vertices.

\begin{figure}[htp]
\begin{center}
\begin{equation*}
DWBC_{high} =
\begin{pmatrix}
\vdots&&&&&&\text{\rotatebox[origin=c]{270}{$\ddots$}}\\
b_{1}&b_{1}&b_{1}&b_{1}&b_{1}&c_{2}&\\
b_{1}&b_{1}&b_{1}&b_{1}&c_{2}&b_{2}&\\
b_{1}&b_{1}&b_{1}&c_{2}&b_{2}&b_{2}&\\
b_{1}&b_{1}&c_{2}&b_{2}&b_{2}&b_{2}&\\
b_{1}&c_{2}&b_{2}&b_{2}&b_{2}&b_{2}&\\
c_{2}&b_{2}&b_{2}&b_{2}&b_{2}&b_{2}&\dotsc\\
\end{pmatrix} =
\begin{array}{c}
\includegraphics[]{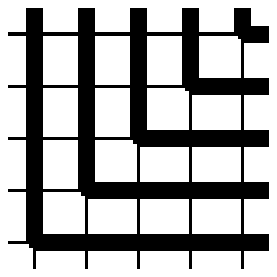}
\end{array}
\end{equation*}
\caption{Domain Wall Boundary Condition High}
\label{high}
\end{center}
\end{figure}

\begin{figure}[htp]
\begin{center}
\begin{equation*}
DWBC_{low} =
\begin{pmatrix}
\ddots&&&&&&&\text{\rotatebox[origin=c]{270}{$\ddots$}}\\
&c_{2}&a_{1}&a_{1}&a_{1}&a_{1}&a_{1}&\\
&a_{2}&c_{2}&a_{1}&a_{1}&a_{1}&a_{1}&\\
&a_{2}&a_{2}&c_{2}&a_{1}&a_{1}&a_{1}&\\
&a_{2}&a_{2}&a_{2}&c_{2}&a_{1}&a_{1}&\\
&a_{2}&a_{2}&a_{2}&a_{2}&c_{2}&a_{1}&\\
&a_{2}&a_{2}&a_{2}&a_{2}&a_{2}&c_{2}&\\
\text{\rotatebox[origin=c]{90}{$\ddots$}}&&&&&&&\ddots\\
\end{pmatrix} =
\begin{array}{c}
\includegraphics[]{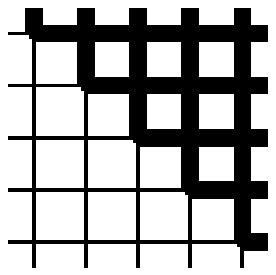}
\end{array}
\end{equation*}
\caption{Domain Wall Boundary Condition Low}
\label{low}
\end{center}
\end{figure}

Domain wall boundary conditions are shown on fig. \ref{high} and
\ref{low}.

For domain wall boundary conditions every path in a 6-vertex
configuration will have one end at the North boundary of the
square and the other end at the East boundary of the square. These
paths can be regarded as level curves of a height function. The
lowest height function is shown on fig. \ref{low} and the highest
height function is shown on fig. \ref{high}.

\subsection{Weights}
The weight of a state is the product of weights of vertices and
the weight of a vertex is determined by rules from fig. \ref{6-v}.
The partition function is sum of weights of all configurations:
\[
Z_{6v}=\sum_{states} \prod_{vertices}w(vertex,state)
\]
where $w(vertex,state)$ is the weight of the vertex (see
fig.\ref{6-v}).

The ratio
\[
\frac{\prod_{vertices}w(vertex,state)}{Z_{6v}}
\]
is the probability of the state. This is the Gibbs measure of the
6-vertex model.

\subsection{Local moves and the graph of states}\label{loc-moves}

Now let us describe local moves in the space of states. Such a move
changes the configuration of arrows at the minimal number of edges
near a given vertex and it acts transitively, i.e. any given state
of the model can be transformed to any other given state of the
model by a sequence of such moves.

\begin{figure}[htp]

\begin{center}
\begin{equation*}
S_{a} =
\begin{pmatrix}
b_{1}& a_{2}\\
c_{2}& b_{2}
\end{pmatrix} =
\begin{array}{c}
\includegraphics[]{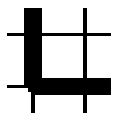}
\end{array}
\end{equation*}
\caption{A local configuration that may be flipped up}
\label{d}
\end{center}
\end{figure}

\begin{figure}[htp]
\begin{center}
\begin{equation*}
S_{b} =
\begin{pmatrix}
c_{2}& c_{1}\\
a_{2}& c_{2}
\end{pmatrix} =
\begin{array}{c}
\includegraphics[]{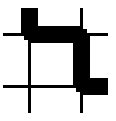}
\end{array}
\end{equation*}
\caption{The local configuration after an up flip has occurred}
\label{u}
\end{center}
\end{figure}

Such moves are most transparent in terms of height functions. There
are two types of local moves:

\begin{itemize}
\item The path from fig. \ref{d} we can move up, i.e. to the path
from fig. \ref{u}. We will call this move flip up.

\item The path from fig. \ref{u} can be moved down, i.e. to the
path from fig. \ref{d}. We will say that this is the flip down.

\end{itemize}

Such moves with all possible surrounding configurations we will
call flips up and down.

For each flippable vertex we introduce effective weight as
follows:
\begin{itemize}

\item For a vertex flippable up the effective weight is the
product of weights of all vertices that can be affected by the
flip, i. e. the vertex itself, and the neighboring vertices to the
North , the East, and the North-East of it.

\item Similarly for a vertex flippable down the effective weight
of it is the product of weights of the vertex itself, and of next
neighboring vertices to the South, West, and South-West of it.

\end{itemize}

The effective weight is always the product of four factors. The
effective weight of vertex $v$ in the configuration $S$ we denote
by $W_v(S)$.
\section{The Markov process}

\subsection{General strategy}
Consider the abstract graph with  vertices being  states of the
model and with edges being local moves. This graph is clearly
connected. Our goal is to construct a random walk on this graph
converging to the probabilistic measure vertices of this graph
which is the Gibbs measure of the 6-vertex model with DW boundary
conditions.

Let us recall some basic facts. Let $\Gamma$ be a finite connected
graph and $q: V(\Gamma)\to {\mathbb R}_+$ be a probabilistic
measure on the set of vertices of $\Gamma$. Let $M=\{p(a\to
b)\}_{a,b\in V(\Gamma)}$ be the matrix of the Markov process
describing a random walk on $\Gamma$. A traveller moves from $a$
to $b$ with the probability $p(a\to b)$.

The matrix $M$ must satisfy the total probability condition:
\[
\sum_bp(a\to b)=1
\]

If in addition it satisfies the detailed balance condition
\[
q(a)p(a\to b)=q(b)p(b \to a)
\]
then it is known that the Markov process converges to $q$. For
details about Markov sampling and estimating convergence times,
see \cite{Sinc}.

Now our goal is to construct such random walk converging to the
Gibbs state of the 6-vertex model. At some point the rate of
convergence of this Markov process becomes an important issue. To
avoid the complicated analysis of estimating mixing times we will
modify the algorithm and will use the ``coupling from the past''
version. This will be explained later.

\subsection{The Markov process for the 6-vertex model}

\paragraph{} We want to construct Markov process which chooses a
vertex at random, then with the probability which we will describe
below it will either flip the configuration up at this vertex, or
will flip it down, or will do nothing. The probability of passing
from the state $S_a$ to the state $S_b$ in this process can be w
written as follows:

\begin{equation}
P(S_{a} \Longrightarrow S_{b}) = \frac{1}{\text{\# vertices}}
\sum_{v}P_{v}(S_{a} \Longrightarrow S_{b})
\end{equation}

\begin{equation}
\begin{split}
= \frac{1}{\text{\# vertices}} \Biggl(&\sum_{(v)non-flip} \delta (S_{a},S_{b})  + \sum_{\substack{(v)flip-up\\ only}}P_{v}(S_{a} \Longrightarrow S_{b}) + \\
&\sum_{\substack{(v)flip-down\\ only}}P_{v}(S_{a} \Longrightarrow
S_{b}) + \sum_{(v)bi-flip}P_{v}(S_{a} \Longrightarrow S_{b})
\Biggr)
\end{split}
\end{equation}

\begin{equation}
\begin{split}
= \frac{1}{\text{\# vertices}} \Biggl(& (\text{\# non-flippable in}S_a)\delta_{S_a,S_b}  + \sum_{\substack{(v)flip-up\\ only}}P_{v}(S_{a} \Longrightarrow S_{b}) + \\
&\sum_{\substack{(v)flip-down\\ only}}P_{v}(S_{a} \Longrightarrow
S_{b}) + \sum_{(v)bi-flip}P_{v}(S_{a} \Longrightarrow S_{b})
\Biggr)
\end{split}
\end{equation}

\begin{equation}
\begin{split}
= \frac{\text{\#non-flippable}}{\text{\# vertices}}\delta_{S_a,S_b} &+ \\
& \frac{\text{\#flippable}}{\text{\#vertices}} \Biggl(\frac{1}{\text{\#flippable}}\sum_{\substack{(v)flip-up\\ only}}P_{v}(S_{a} \Longrightarrow S_{b}) + \\
& \frac{1}{\text{\#flippable}}\sum_{\substack{(v)flip-down\\
only}}P_{v}(S_{a} \Longrightarrow S_{b}) +
\frac{1}{\text{\#flippable}}\sum_{(v)bi-flip}P_{v}(S_{a}
\Longrightarrow S_{b}) \Biggr)
\end{split}
\end{equation}
Here $\text{\#flippable}$ is the number of flippable vertices and
$\text{\#verices}$ is the total number of vertices.

Algorithmically, this means that we do the following:
\begin{enumerate}
\item With probability
$P=\frac{\text{\#non-flippable}}{\text{\#vertices}}$, do nothing
(that is, restart the loop.) \item With probability
$P=\frac{\text{\#flip-up-only + \#flip-down-only +
\#bi-flip}}{\text{\#vertices}}=\frac{\text{\#flippable}}{\text{\#vertices}}$,
continue to the next part.
\end{enumerate}
If the algorithm continues, select a flippable vertex with the
probability:
\begin{equation}
P(selection) = \frac{1}{\#flippables}
\end{equation}
At this selected vertex the configuration can be either flippable
only up, or only down, or in both directions. Depending on this
proceed according to the following rules:

Three possible conditions now exist:
\begin{enumerate}

\item The vertex is flippable down only.  Two options:
\begin{itemize}
\item Flip vertex down with probability $P_v(S_{a} \Longrightarrow
S_{b}) =\rho W_v(S_{b})$ \item Stay with probability $P_v(stay) =
1 - \rho W_v(S_{b})$
\end{itemize}

\item The vertex is flippable up only. Two options:
\begin{itemize}
\item Flip vertex up with probability $P_v(S_{a} \Longrightarrow
S_{b}) =\rho W_v(S_{b})$ \item Stay with probability $P_v(stay) =
1 - \rho W_v(S_{b})$
\end{itemize}

\item The vertex is flippable up and down. Three options:
\begin{itemize}
\item Flip vertex down with probability $P_v(S_{a} \Longrightarrow
S_{b}) =\rho W_v(S_{b})$ \item Flip vertex up with probability
$P_v(S_{a} \Longrightarrow S_{b^{'}}) =\rho W_v(S_{b^{'}})$ \item
Stay with probability $P_v(stay) = 1 - \rho W_v(S_{b}) - \rho
W_v(S_{b^{'}})$
\end{itemize}
\end{enumerate}

Here $W_v(S_{b^{'}})$ $W_v(S_{b})$ are the effective weights of
the vertex $v$ in the states obtained by flipping up or down at
this vertex from the state $S_{a}$. Effective weights were
described in section \ref{loc-moves}.

The parameter $\rho$ is chosen such that all probabilities of
transitions should be positive. In other words it should satisfy
all conditions $\rho<\frac{1}{W_v(S')}$where $v$ is a vertex
flippable in the state $S$ either up or down, but not biflippable
and $S'$ is the configuration after the flip. At every biflippable
vertex in the state $S$ we should have
$\rho<\frac{1}{W_v(S')+W(S'')}$ where $S'$ is the result of the
flipping $S$ up at $v$ and $S''$ is the result of the flip down.

This process satisfies the detailed balance condition, and the
total probability condition. Since the graph of states with edges
being local moves is connected, this process converges to the
Gibbs state of the 6-vertex model. The process also depends on the
choice of $\rho$. It slows down when $\rho$ is small.

\section{Random states in the 6-vertex model with DW boundary
conditions}\label{results}
\subsection{Phases in the 6-vertex model}
\subsubsection{}

One can write weights of the 6-vertex model as
\begin{eqnarray}\label{w-magn}
a_1&=&\exp(-\frac{E_1-H_x-H_y}{T}),\ a_2=\exp(-\frac{E_1+H_x+H_y}{T}) \, \\
b_1&=&\exp(-\frac{E_2+H_x-H_y}{T}) , \
b_2=\exp(-\frac{E_2-H_x+H_y}{T}) \, \\
c_1&=&c_2=\exp(-\frac{E_3}{T})  \,
\end{eqnarray}
Here $E_1, E_2$, and $E_3$ are energies of the interaction of
arrows (or energies associated with the local shape of level
curves of the height function) and $H_x$ and $H_y$ are magnetic
fields.

In this interpretation arrows can be regarded as spins interacting
with the magnetic filed such that the energy of a vertical arrow
is $\pm H_x$ depending on whether the arrow is heading up or down.
The energy of a horizontal arrow is $\pm H_y$ depending on whether
it is oriented left or right. We assigned the energy of an arrow
to the energies of adjacent vertices.

Notice that since the total number of $c_1$- and $c_2$-vertices
satisfy the relation $n(c_1)-n(c_2)=N$ the partition function
changes by an overall factor only when we change $c_1/{c_2}$. The
total numbers of $a$ and $b$ vertices satisfy similar relations:
$n(a_1)=n(a_2)$ and $n(b_1)=n(b_2)$. Because of this for the
square lattice with DW boundary conditions we can set $a_1=a_2=a$,
$b_1=b_2=b$, and $c_1=c_2=c$ without loosing generality.

\subsubsection{}Let us recall the phase diagram of the 6-vertex
model \cite{LWu}, \cite{Bax} with periodic boundary conditions in
the absence of magnetic fields. The important characteristic of
the model is the parameter
\[
\Delta=\frac{a^2+b^2-c^2}{2ab}
\]

\begin{figure}[htp]
\begin{center}
\includegraphics[width=1.5in]{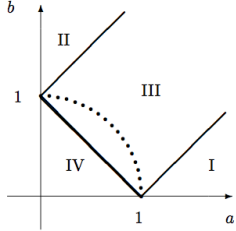}
\caption{The phase diagram for the six-vertex model in terms of the
  weights of $a$ and $b$,
assuming $c=1$.}
\label{phases}
\end{center}
\end{figure}

The phase diagram for the 6-vertex model with periodic boundary
conditions in the
absence of magnetic fields is shown on fig. \ref{phases}.

\begin{figure}[htp]
\begin{center}
\includegraphics[]{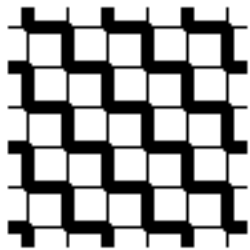}
\caption{Antisegnetoelectric phase - in this phase, zig-zag paths alternate with zig-zags formed by empty edges}
\label{ase-phase}
\end{center}
\end{figure}

There are four phases:
\begin{enumerate}
\item Phase I: $a>b+c$($\Delta>1$). This an ordered phase where
there are two possibilities for the ground state. It either
consists of $a_1$-vertices or of $a_2$-vertices. In either case
any change in the ground state gives the state with the total
number of $b$ and $c$ vertices comparable with the linear size $N$
of the system. Thus, as $N\to \infty$ the energy of these two
ground states is macroscopically  separated from the energy of
other states. In other words these are two frozen ground states.

\item Phase II: $b>a+c$ ($\Delta>1$). This is an ordered phase
with double degeneracy of the ground state. The first possibility
is when all vertices are $b_1$ vertices, the second possibility is
when all vertices are $b_2$-vertices. As in case of phase I, this
is a frozen phase.

\item Phase III: $a,b,c<\frac{a+b+c}{2}$($ |\Delta|<1$). This is a
disordered phase. Local correlation functions decay as a power of
the distance in this phase. These are the values of $a,b,c$ when
$|\Delta|<1$. In particular, the free fermionic curve $\Delta=0$
lies entirely in this phase. It is shown by the dotted segment of
the circle on fig. \ref{phases}.

\item Phase IV: $c>a+b$ ($\Delta<-1$). This is an ordered phase
with so-called antisegnetoellectic ordering (see fig.
\ref{ase-phase}). The ground state in this case consists of
alternating $c_1$ and $c_2$ vertices. It is double degenerate due
to the breaking of $Z_2$-translational symmetry. In this case
microscopic  deviations from the ground state are possible. There
is a finite correlation length in the system and local correlation
functions decay exponentially.

\end{enumerate}

For details about phase transitions, magnetization, and the
antiferroelectric phase etc. see \cite{LWu} and \cite{Bax}.

\subsection{ The structure of a random state}
\subsubsection{Free fermionic point} This is the case when
$\Delta=0$.  It is convenient to parameterize
weights in this case as
\[
a=\rho \cos u , \  b=\rho\sin u, \ c=\rho.
\]
When $a=b=1/{\sqrt{2}}$ this model is equivalent to the domino tiling of
the Aztec diamond. The limit shape was computed analytically in
\cite{AztDiam} and is a circle.

\begin{figure}[!htp]
\begin{center}
\begin{tabular}{cccc}
\includegraphics[width=1.5in]{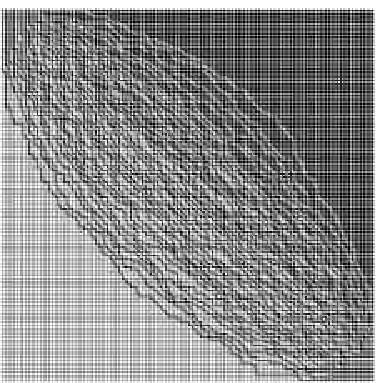}
&\includegraphics[width=1.5in]{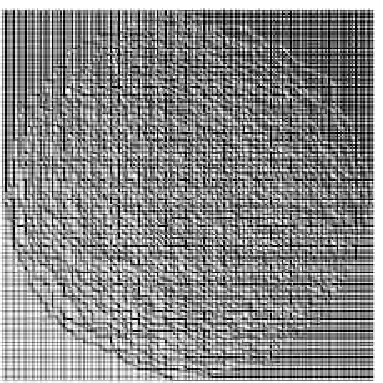}
&\includegraphics[width=1.5in]{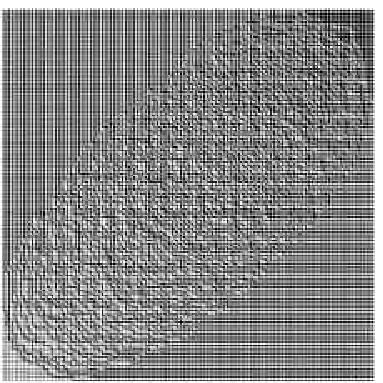}\\
$2a=b$&$a=b$&$a=2b$\\
\end{tabular}
\caption{Free fermionic point with $\Delta=0$}
\label{delta0free}
\end{center}
\end{figure}

The height functions of the average states for several values of
the parameter $u$ are shown on fig. \ref{delta0free}.  For
$\Delta=0$, the limit shapes can be computed explicitly using
methods of \cite{Ken-Okoun-Shef} and they are ellipses, which
agrees with fig.  \ref{delta0free}.

\subsubsection{Ordered phases }
In phase I the $a$-vertices dominate and the Gibbs state in this
case is given by the lowest height function fig. \ref{low}.

In phase II the $b$ vertices dominate and according to \cite{LWu}
we should expect that the average state will be the state with the
domination of $b$-vertices. In other words the average state in
this case is given by the highest height function fig. \ref{high}.

\subsubsection{Disordered phase }

In this case it is convenient to use the following parametrization
of weights:

\[
a=r\sin (\gamma-u), \ b=r\sin u, \ c=r\sin \gamma
\]
with $0<\gamma<\pi$, $0<u<\gamma$, and $r>0$. In this
parametrization $\Delta=\cos \gamma$.

Phase III contains the free fermionic curve $\Delta=0$. Since all
this phase is critical one may expect that the nature of the Gibbs
states will be similar for all parameters $a,b,c$ in this region.
In particular, one can expect the existence of the limit shape as
in the case $\Delta=0$. The particular form of the limit shape may
vary but the following common features should common for all
values of $a,b,c$ in this region:
\begin{itemize}
\item The limit shape is a smooth curve having exactly one common
point with each side of the square. At this point the limit shape
is tangent to the side of the square.

\item Inside of the boundary of the limit shape the height
function is a smooth function and it has continuous first
derivative at the boundary. The second derivative has a
discontinuity in the normal direction to the boundary of the limit
shape.

\item Outside of the boundary of the limit shape the height
function is linear.

\end{itemize}

\begin{figure}[!htp]
\begin{center}
\begin{tabular}{cccc}
\includegraphics[width=1.5in]{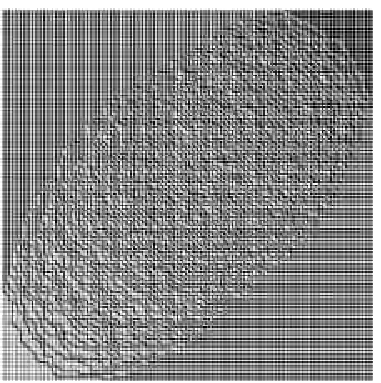}
&\includegraphics[width=1.5in]{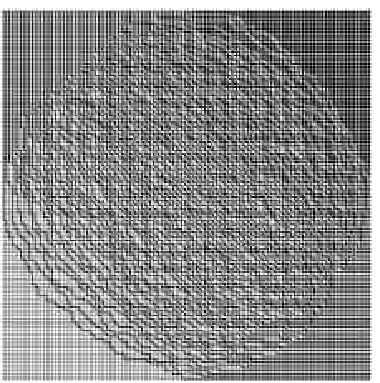}
&\includegraphics[width=1.5in]{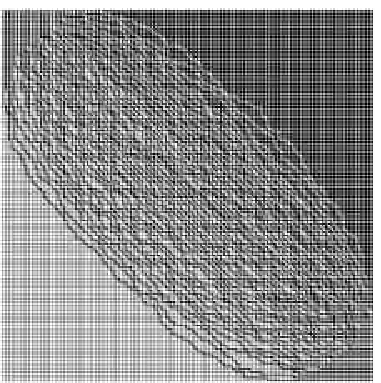}\\
$2a=b$&$a=b$&$a=2b$\\
\end{tabular}
\caption{Disordered phase with $\gamma=\frac{\pi}{4}$}
\label{disordered-deltapi4}
\end{center}
\end{figure}

\begin{figure}[h]
\begin{center}
\begin{tabular}{cccc}
\includegraphics[width=1.5in]{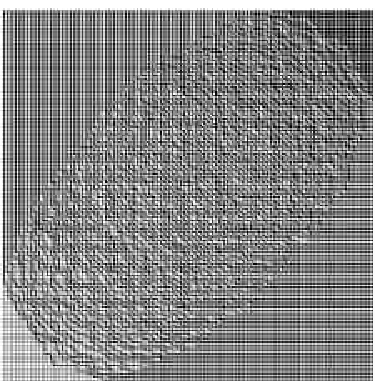}
&\includegraphics[width=1.5in]{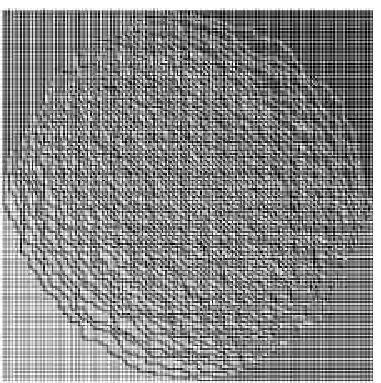}
&\includegraphics[width=1.5in]{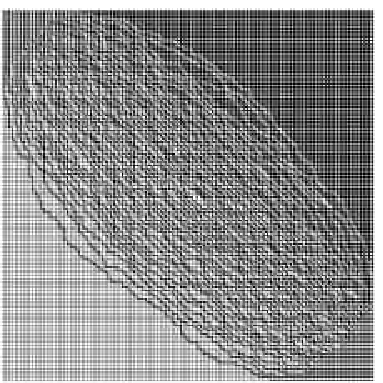}\\
$2a=b$&$a=b$&$a=2b$\\
\end{tabular}
\caption{Disordered phase with $\gamma=\frac{\pi}{5}$}
\label{disordered-deltapi5}
\end{center}
\end{figure}

\begin{figure}[!htp]
\begin{center}
\begin{tabular}{cccc}
\includegraphics[width=1.5in]{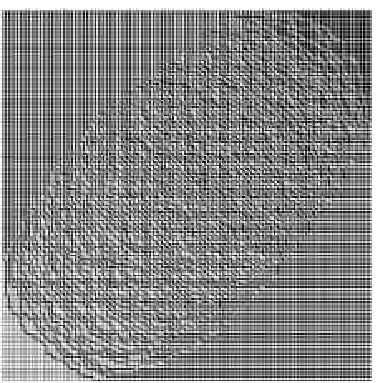}
&\includegraphics[width=1.5in]{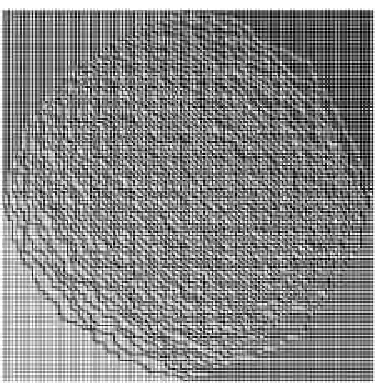}
&\includegraphics[width=1.5in]{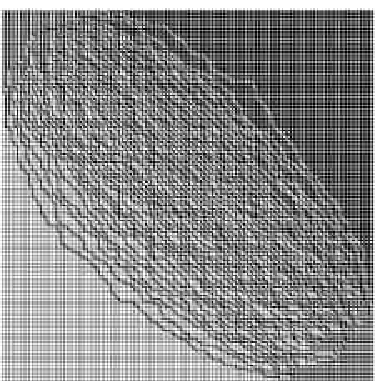}\\
$2a=b$&$a=b$&$a=2b$\\
\end{tabular}
\caption{Disordered phase with $\gamma=\frac{\pi}{8}$}
\label{disordered-deltapi8}
\end{center}
\end{figure}

Examples of Gibbs states in the disordered phase are shown on fig.
\ref{disordered-deltapi4}, \ref{disordered-deltapi5},
\ref{disordered-deltapi8}.

\subsubsection{The antiferroelectric phase }
\begin{figure}[!htp]
\begin{center}
\begin{tabular}{cccc}
\includegraphics[width=1.5in]{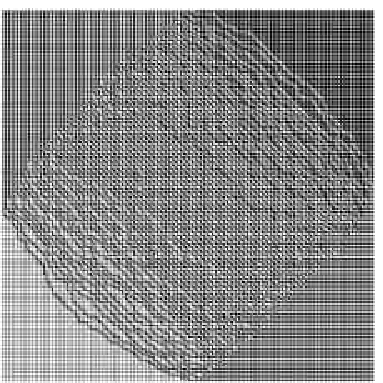}
&\includegraphics[width=1.5in]{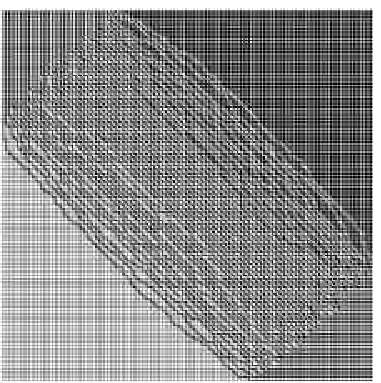}
&\includegraphics[width=1.5in]{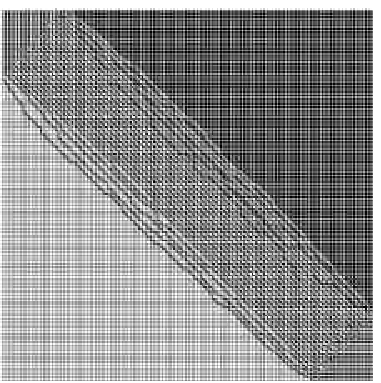}\\
$2a=b$&$a=b$&$a=2b$\\
\end{tabular}
\caption{Antiferroelectric phase with $\Delta=-3$}
\label{c-domination}
\end{center}
\end{figure}

This region $c>a+b$ is the one which is non-critical and which is
also not ordered. In the periodic case the ground state has the
domination of $c$-vertices as it is shown on fig.
\ref{c-domination}.

It is convenient to use the parameterization

\[
a=r \sinh(\eta-u), \ b=r \sinh u, \ c=r \sinh \eta
\]
with $0<u<\eta$. In this parameterization  $\Delta=-\cosh \eta$.

In the case of DW boundary conditions there is a competition
between very rigid restrictions on the states near the boundary
which allows only $a$ and $b$ vertices near the boundary and the
tendency of the system to have as much as possible of $c$
vertices.

Numerical simulations show that these competing tendencies resolve
in the separation of three phases. It is fairly convincing from the fig.
\ref{c-domination} that the following should take place:
\begin{itemize}
\item The system forms a macroscopical droplet of the
antiferroelectric phase with the boundary that does not touch the
square. The height function in this domain is linear. The boundary
of this domain has four cusps pointing towards sides of the square
lattice.  This phase is noncritical. Correlation functions in this region decay exponentially.

\item Near the boundary the system is ordered. This ordered region
is bordered by the disordered region where the height function is
smooth. The disordered phase is critical.  There is a finite magnetization, which means there are excitations with linear dispersions and therefore correlation functions decay according to a power law. The boundary between ordered and disordered phases is a
smooth curve with the features similar to the $|\Delta|<1$ case.

\end{itemize}

\begin{figure}[!htp]
\begin{center}
\begin{tabular}{cccc}
\includegraphics[width=1.5in]{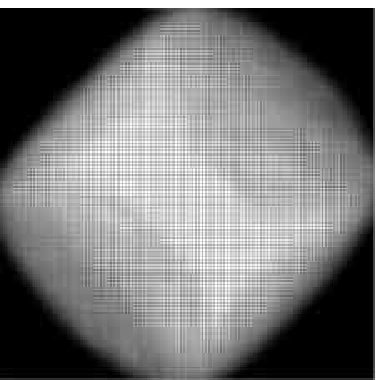}
&\includegraphics[width=1.5in]{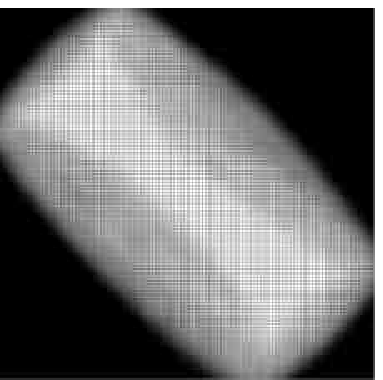}
&\includegraphics[width=1.5in]{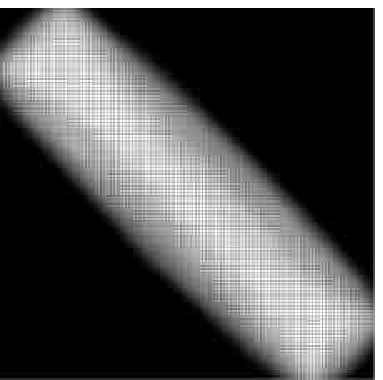}\\
$2a=b$&$a=b$&$a=2b$\\
\end{tabular}
\caption{c-density plots of the antiferroelectric phase with
$\Delta=-3$} \label{c-dominationcden}
\end{center}
\end{figure}

\section{Conclusion}

We demonstrated that local Markov sampling  for the 6-vertex model
with domain wall boundary conditions indicates that the system
develops a macroscopical droplet of $c$-vertices when $\Delta<-1$.
For these computations it is not essential that the ground state
of the 6-vertex model in this phase is doubly degenerate. This
degeneracy corresponds to the translation by one step in the
North-East direction on fig. \ref{c-domination}.

This degeneracy is important in the computation of correlation
functions and other observables. The two ground states correspond
to the two parts of the graph of states which are connected by a
"very narrow neck" in the limit $N\to \infty$. We will address
this problem in the next publication. The existence of the droplet
can be seen from results of \cite{SZ} where a different numerical
method was used. It would be interesting to compare the methods.

The droplet of $c$-vertices is similar to the facets in dimer
models. The shape of the droplet and of the surrounding critical
phase is similar to the corresponding shapes in the dimer model on
the square-octagon lattice \cite{Ken-Okoun-Shef}.

The local Markov sampling which we used here is equally effective
for other boundary conditions in the 6-vertex model. Some of the
results for more complicated boundary conditions can be found in
\cite{Er}.

\appendix

\section{Functions and Implementations}

\subsection{Main loop}
\paragraph{}The following tasks must be completed by the Main loop function:

\begin{enumerate}
\item Import the matrix from a text file \item Build flippables
list \item Set weights \item Define $\rho$:

\begin{equation}
\rho = \frac{1}{\max \{ \text{\emph{weight combinations for all
flip types}} \} }
\end{equation}

\item Loop the following actions, and after a certain defined
number of successful flips, output a file with the current matrix
(and status of the Markov Chain) in it.

\begin{enumerate}

\item Generate a random real, $rand$,  between 0 and 1
\begin{enumerate}
\item If $\frac{\text{\#flip-up-only + \#flip-down-only +
\#bi-flip}}{\text{\#vertices}} \geq rand$, continue to (b). \item
Otherwise, go to (a).
\end{enumerate}

\item Select a random flippable position with probability \newline
$P(selection) = \frac{1}{\text{\#flip-up-only + \#flip-down-only +
\#bi-flip}}$ \newline by calling the Get Flippable Position
function. \item Call Get Weight (which is now scaled by the value
of $\rho$, to ensure that it always returns a value less than 1)
to get the probability of an up flip and/or a downflip at the
flippable location chosen.

\item Generate a random real, $rand$, between 0 and 1.
\begin{enumerate}
\item For up or down-only flips, iff $W(S_{b}) \geq rand$, execute
the flip by calling the Execute Flip function, else restart main
loop. \item For positions that can flip up and down, iff $W(S_{b})
\geq rand$, execute the flip corresponding to $S_{b}$, else iff
$W(S_{b^{'}}) \geq rand$, execute the flip corresponding the
$S_{b^{'}}$, else restart main loop. In practice, this means that
once a vertex which can be flipped either way is chosen, simply
divide up the probabilities of each flip occurring as discussed
earlier.
\end{enumerate}

\end{enumerate}

\end{enumerate}

\subsection{Execute Flip}
\begin{enumerate}
\item If type is high
\begin{enumerate}
\item Change the entry in the list of Flippables for the vertex
chosen to make a high flip impossible.

\item Define the following positions:
\begin{enumerate}
\item One = the original position = Base
\item Two = (+1, +0) = Right
\item Three = (+1, +1) = Up Right
\item Four = (+0, +1) = Up
\item Left = (-1, +0)
\item Down=(+0, -1)
\item UpLeft = (-1, +1)
\item UpRight = (+1, +1)
\item DownLeft = (-1, -1)
\item DownRight = (+1, -1)
\end{enumerate}

\item Replace 4 parts
\begin{enumerate}
\item Set Contents of Position One = FlipToOne (Position One, High)
\item Set Contents of Position Two = FlipToTwo (Position Two, High)
\item Set Contents of Position Three = FlipToThree (Position Three, High)
\item Set Contents of Position Four = FlipToFour (Position Four, High)
\end{enumerate}

\item If Up, Down, Right, or Left Positions become flippable (call
Get Is Flippable on each to check), add them to the Flippables
List. \item Call Fix Low End.

\end{enumerate}

\item If type is low
\begin{enumerate}

\item Change the entry in the list of Flippables for the entry
chosen to make a low flip impossible.

\item Define the following positions:
\begin{enumerate}
\item One = (-1, -1) = Down Left
\item Two = (+0, -1) = Down
\item Three = the original position = Base
\item Four = (-1, +0) = Left
\item Right = (+1, +0)
\item Up = (+0, +1)
\item UpLeft = (-1, +1)
\item UpRight = (+1, +1)
\item DownLeft = (-1, -1)
\item DownRight = (+1, -1)
\end{enumerate}

\item Replace 4 parts
\begin{enumerate}
\item Set Contents of Position One = FlipToOne (Position One, Low)
\item Set Contents of Position Two = FlipToTwo (Position Two, Low)
\item Set Contents of Position Three = FlipToThree (Position
Three, Low) \item Set Contents of Position Four = FlipToFour
(Position Four, Low)
\end{enumerate}

\item If Up, Down, Right, or Left Positions become flippable (call
Get Is Flippable on each to check), add them to the Flippables
List. \item Call Fix High End.

\end{enumerate}
\end{enumerate}

\subsubsection{Fix High End}
\begin{enumerate}

\item Define the following positions:
\begin{enumerate}
\item HighCreateDownLeft = (-1, -1) \item HighDeleteDown = (+0,
-1) \item HighDeleteDownDownLeft = (-1, -2) \item
HighDeleteDownRighLeft = (-2, -1) \item HighDeleteLeft = (-1, +0)
\end{enumerate}

\item If HighCreateDownLeft is flippable, add it to the High
Flippables List.
\item Delete 4 potential flippables on the High
End.
\item If any of HighDeleteDown, HighDeleteDownDownLeft,
HighDeleteDownLeftLeft, or HighDeleteLeft exists in the High
Flippables list, remove them from the list.

\end{enumerate}

\subsubsection{Fix Low End}

\begin{enumerate}

\item Define the following positions:
\begin{enumerate}
\item LowCreateUpRight = (+1, +1) \item LowDeleteUp = (+0, +1)
\item LowDeleteUpUpRight = (+1, +2) \item LowDeleteUpRighRight =
(+2, +1) \item LowDeleteRight = (+1, +0)
\end{enumerate}

\item If LowCreateUpRight is flippable, add it to the Low
Flippables List.
\item Delete 4 potential flippables on the Low End.
\item If any of LowDeleteUp, LowDeleteUpUpRight,
LowDeleteUpRightRight, or LowDeleteRight exists in the Low
Flippables list, remove them from the list.
\end{enumerate}

\subsection{Get Weight}
\begin{enumerate}
\item Get contents of four surrounding positions if the flip
occurred.
\item Multiply weights together corresponding to the
contents of the four positions.
\item Multiply (new weight configuration product) * $\rho$
\end{enumerate}

\subsection{Flip To}
\paragraph{} FlipTo functions take a position and a type and return what the vertex at the given position would be after the flip of the type specified.

\begin{enumerate}
\item FlipToOne (Position, Type)
\begin{enumerate}
\item If type is high: \newline If vertex was $a_{1}$, it will be
$c_{1}$; if it was $c_{2}$, it will be $a_{2}$ \item If type is
low: \newline If vertex was $c_{1}$, it will be $a_{1}$; if it was
$a_{2}$, it will be $c_{2}$
\end{enumerate}

\item FlipToTwo
\begin{enumerate}
\item If type is high: \newline If vertex was $b_{2}$, it will be
$c_{2}$; if it was $c_{1}$, it will be $b_{1}$ \item If type is
low: \newline If vertex was $c_{2}$, it will be $b_{2}$; if it was
$b_{1}$, it will be $c_{1}$
\end{enumerate}
\item FlipToThree
\begin{enumerate}
\item If type is high: \newline If vertex was $a_{2}$, it will be
$c_{1}$; if it was $c_{2}$, it will be $a_{1}$ \item If type is
low: \newline If vertex was $c_{1}$, it will be $a_{2}$; if it was
$a_{1}$, it will be $c_{2}$
\end{enumerate}

\item FlipToFour
\begin{enumerate}
\item If type is high: \newline If vertex was $b_{1}$, it will be
$c_{2}$; if it was $c_{1}$, it will be $b_{2}$ \item If type is
low: \newline If vertex was $c_{2}$, it will be $b_{1}$; if it was
$b_{2}$, it will be $c_{1}$
\end{enumerate}
\end{enumerate}

\subsection{Get Flip Position}
\begin{enumerate}
\item Generate a random integer between 1 and the total number of
flippable positions; that is, the number of up-flip only plus the
number of down-flip only plus the number of bi-flips. \item Choose
the corresponding element in the Flippable Positions list to the
random number chosen.
\end{enumerate}

\subsection{Get Is Flippable}
\paragraph{}
Get Is Flippable should check the status of a position to
determine if it is flippable.

\begin{enumerate}
\item High flippables must be $a_{1}$ or $c_{2}$ vertices and must
have empty upper right corners (upper right corner must be $a_{2}$
or $c_{2}$).  High flippable positions must have an x axis
coordinate that is less than or equal to the width of the matrix Ð
1 (where 0,0 is the origin) and a y axis coordinate that is less
than or equal to the height of the matrix Ð1. \item Low flippables
must be $a_{1}$ or $c_{1}$ vertices and must have empty lower left
corners (lower left corner must be $a_{2}$ or $c_{1}$).  Low
flippable positions must have an x axis coordinate that is no less
than 1 (where 0,0 is the origin), and a y axis coordinate that is
no less than 1.
\end{enumerate}

\newpage
\section{Images of the $N=1000$ matrix}

\begin{figure}[!htp]
\begin{center}
\includegraphics[width=5in]{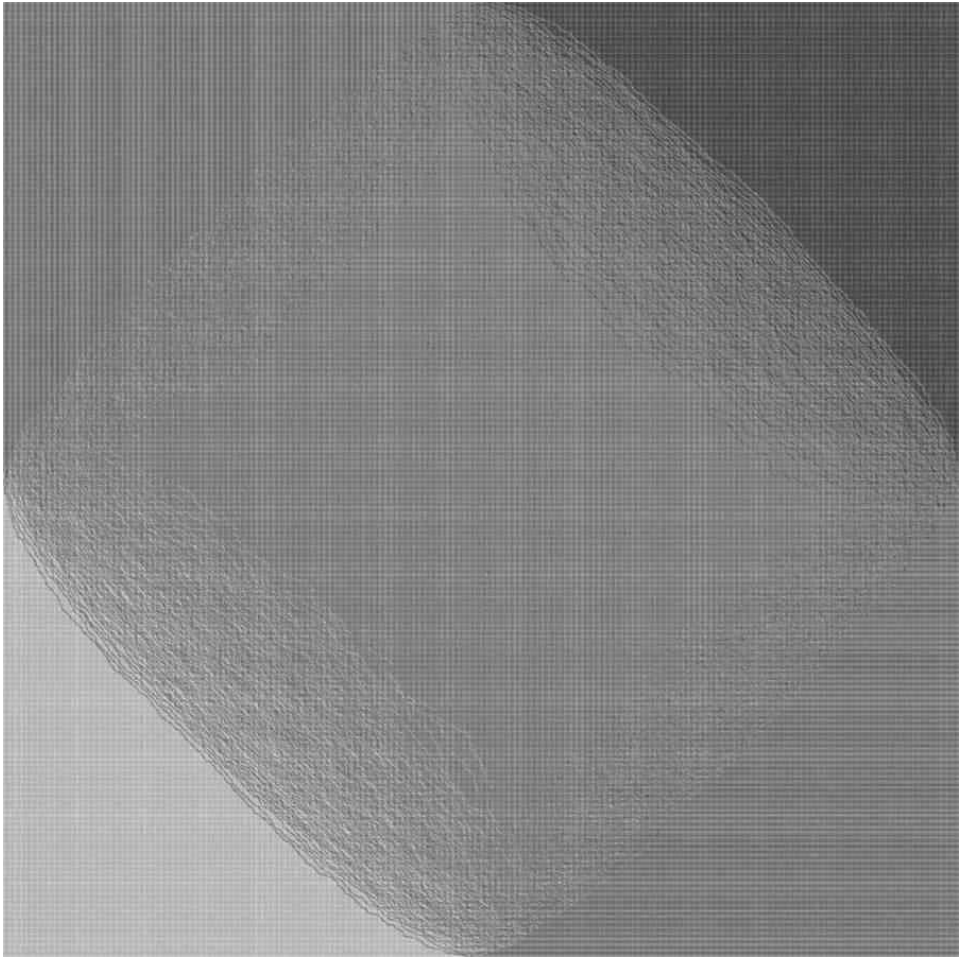}
\caption{$N=1000$ plot for the antiferroelectric phase with $\Delta = 3$ and $2a = b$}
\label{1k}
\end{center}
\end{figure}
\newpage

\begin{figure}[!htp]
\begin{center}
\includegraphics[width=5in]{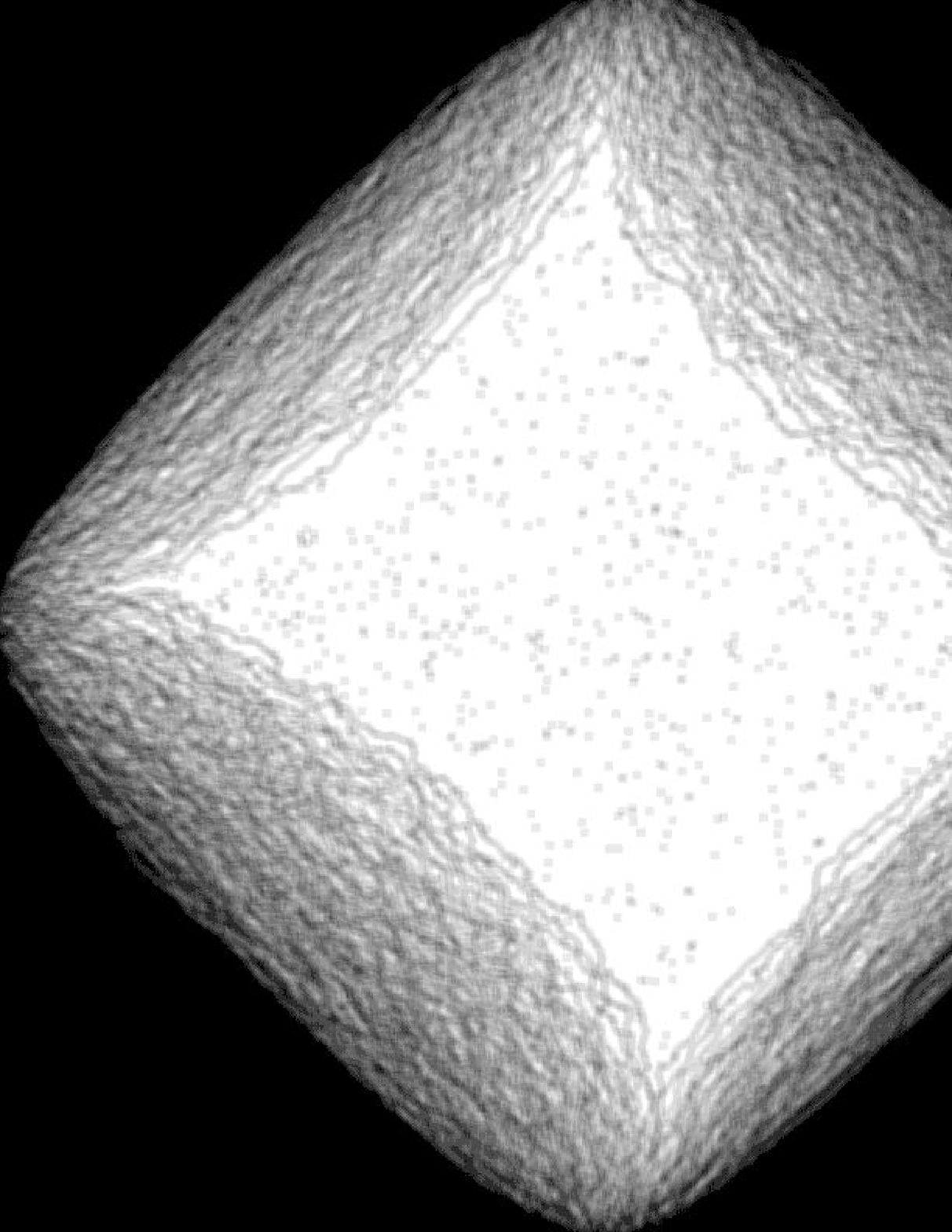}
\caption{$N=1000$ $c$-vertex density plot for the antiferroelectric phase with $\Delta = 3$ and $2a = b$}
\label{1kcden}
\end{center}
\end{figure}

\end{document}